# Guided Surface Plasmon Mode of Semicircular Cross Section Silver Nanoridges

**Junpeng Guo[*] and Zeyu Pan**

*Department of Electrical and Computer Engineering, University of Alabama in Huntsville,*

*Huntsville, Alabama 35899, USA*

*Corresponding author: guoj@uah.edu

Tightly confined plasmon waveguide modes supported by semicircular cross section top silver nanoridges are investigated in this paper. Mode field profiles, dispersion curves, propagation distances, confinement factors, and figure-of-merits of semicircular top silver nanoridge plasmon waveguide mode are calculated for different radii of curvature at different wavelengths. It is found that semicircular top silver nanoridges support tightly confined quasi-TEM plasmon waveguide modes. Semicircular top silver nanoridge mode has longer propagation distance and higher figure-of-merit than that of the cylindrical silver nanowire of the same radius of curvature.

## 1. Introduction

Surface plasmons are free electron density oscillations on the metal-dielectric interfaces, and can propagate along the metal-dielectric boundaries in the form of surface plasmon polariton (SPP) waves with highly confined plasmon optical waveguide modes [1, 2]. Surface plasmon modes in various waveguide structures, such as thin metal films [3-7], finite width thin film metal stripes [8-14], metal-dielectric layers [15-20], trenches in the metal surfaces [21-27], metal wedges [26-32], dielectric-loaded metal films [33-37], and cylindrical metal wires [38-43] have been extensively investigated in the past. The quest for strong mode confinement and low propagation attenuation has motivated the research efforts on various waveguide structures. Recently, well confined SPP modes of flat-top and triangular cross section silver nanoridges have been investigated [44, 45]. Round-top gold nanoridge surface plasmon waveguides have been fabricated with the focused ion beam (FIB) milling technique [32].

This paper presents comprehensive numerical investigations on the plasmon waveguide mode supported by semicircular cross section top silver nanoridges. The mode field profiles, mode indices, dispersion curves, propagation distances, mode sizes, confinement factors, and figure-of-merits of these silver nanoridges with different semicircular radii are calculated and presented in this paper. The metal nanoridge mode properties are compared with the plasmon waveguide mode of the cylindrical metal nanowire of the same radius. Semicircular top metal nanoridges provide more uniform surface plasmon mode energy distribution than that of flat-top and triangular cross section metal nanoridge waveguides [44, 45]. This feature is important in applications that require expanded surface interaction areas such as in chemical and biological sensing. Semicircular top nanoridge waveguides can be fabricated with the focused ion beam (FIB) milling technique, which has been shown to be a powerful nanofabrication technique for realizing nanoscale structures and shapes in various materials [46].

## 2. SEMICIRCULAR CROSS SECTION SILVER NANORIDGE PLASMON WAVEGUIDES

Figure 1(a) and (b) are the three-dimensional (3D) view and the cross section of a semicircular metal nanoridge with the semicircular radius ($r$). The metal nanoridge plasmon waveguide is extended in the $z$ direction with a ridge width of $2r$ in the $x$ direction. The plasmon waveguide mode propagates along the ridge top in the *z* direction. The geometry of the semicircular cross section top nanoridge is symmetric in the *x* dimension, and non-symmetric in the *y* dimension. We assume that the height of the metal ridge is sufficiently high that the existence of the substrate does not influence the ridge mode.

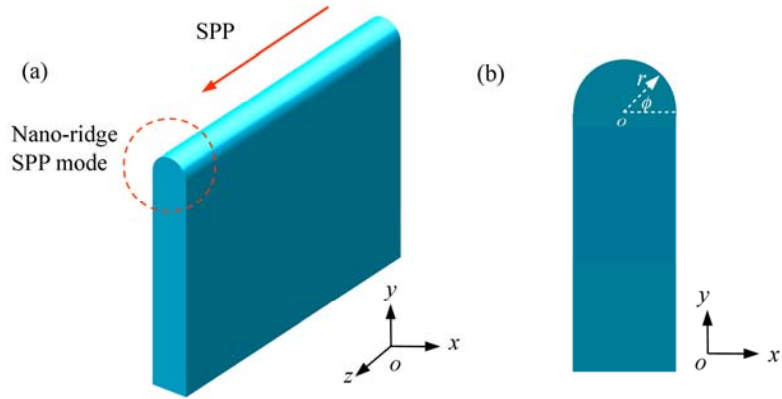

Fig. 1. (a) 3D view of the semicircular cross section nanoridge plasmon waveguide. (b) Cross section of the semicircular cross section nanoridge waveguide.

We use a full-vectoral finite-difference mode solver [47] (*Lumerical Solutions, Inc.*) to calculate the plasmon waveguide modes of the silver nanoridges. The finite-difference method uses Yee's 2D mesh with appropriate boundary conditions to numerically solve Maxwell's equations in the frequency domain. We consider the situation where the surrounding medium of the nanoridges is air ($\varepsilon_d = 1.0$), and the nanoridges are made of silver with its electric permittivity $\varepsilon_m = -127.5 - 5.3j$ at the wavelength of 1.55 $\mu m$ [48]. For the semicircular ridge with a radius of 250 *nm*, the mode index and the attenuation coefficient are found to be $n_{eff}=1.012-0.00044j$ and 154.74 *dB/cm*, respectively. Although this study investigates silver as the nanoridge material, the analysis can obviously be extended to other types of metals and high electron density materials, such as heavily doped semiconductors [49-51] at different wavelengths.

We calculated the real and imaginary parts of the mode field components of the 250 *nm* radius semicircular silver nanoridge plasmon waveguide at the 1.55 *μm* wavelength. It is found that the real parts of transverse components: $E_r$, $E_\phi$, $H_r$, $H_\phi$ are three orders of magnitude larger than their corresponding imaginary parts. But the real parts of longitudinal components $E_z$, $H_z$ are two orders of magnitude less than the corresponding imaginary parts. In the polar coordinate system, we plot the real part of $E_r$, the real part of $E_\phi$, and the imaginary part of $E_z$ mode profiles in the Fig. 2(a-c). Also, we plot the real part of $H_r$, the real part of $H_\phi$, and the imaginary part of $H_z$ mode profiles in the Fig. 2(d-f). The major component of the electric field is the transverse $E_r$ component. The longitudinal component of the electric field $E_z$ is about one order of magnitude less than the transverse $E_r$ component. The major magnetic field component is the transverse $H_\phi$ component, while the longitudinal component of the magnetic field $H_z$ is also one order of magnitude less than the transverse $H_\phi$ component. Based on these results, the semicircular cross section nanoridge plasmon modes can be considered as a quasi-circular electromagnetic mode. From Fig. 2, it can be seen that the $E_r$ mode profile and $H_\phi$ mode profile are symmetrical with respect to the center of the metal ridge (*x*=0 plane), while the $E_\phi$ and $H_r$ profiles are anti-symmetrical with respect to the *x*=0 plane.

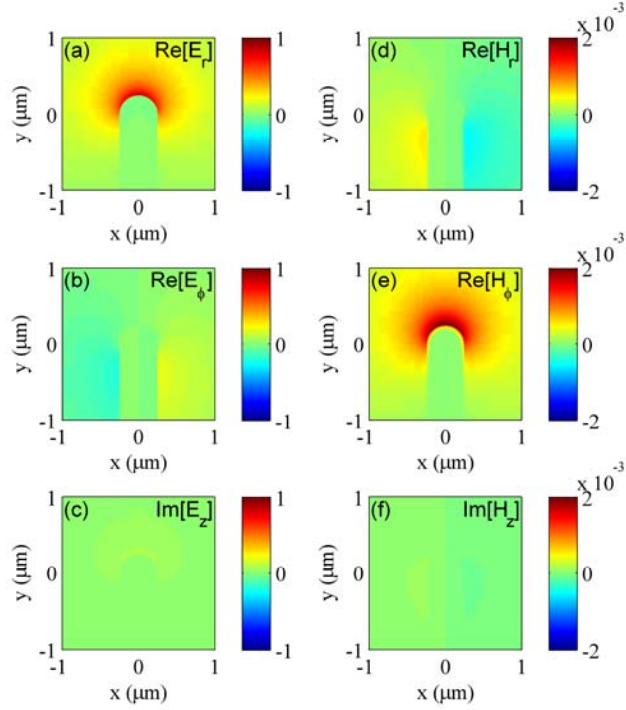

Fig. 2. The mode profiles of a semicircular silver nanoridge with $r = 250$ *nm* at 1.55 $\mu m$ wavelength: (a) real part of the electric field $E_r$, (b) real part of the electric field $E_\phi$, (c) imaginary part of the electric field $E_z$, (d) real part of the magnetic field $H_r$, (e) real part of the magnetic field $H_\phi$, (f) imaginary part of the magnetic field $H_z$.

To compare with the surface plasmon waveguide mode of the cylindrical silver nanowire, we calculated the mode profiles of the real part of $E_r$, real part of $E_\phi$, imaginary part of $E_z$, real part of $H_r$, real part of $H_\phi$, and imaginary part of $H_z$ for the 250 *nm* radius cylindrical silver nanowire at the 1.55 $\mu m$ wavelength. The mode profiles are shown in the Fig. 3(a-f). The major component of the electric field is the transverse component $E_r$, as expected. The longitudinal component of the electric field $E_z$ is seen to be much smaller than the transverse component $E_r$. The major magnetic field component is the transverse component $H_\phi$, while the longitudinal component of the magnetic field $H_z$ is also seen to be much smaller than the transverse component. Therefore, the semicircular top nanoridge plasmon waveguide mode can be considered as a quasi-cylindrical plasmon waveguide mode.

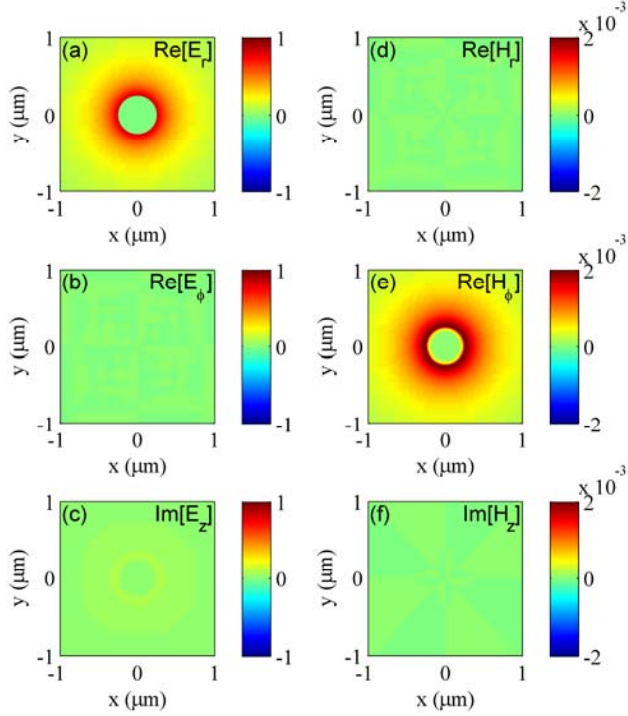

Fig. 3. The mode profiles of a cylindrical plasmon waveguide with $r = 250$ $nm$ at 1.55 $\mu m$ wavelength: (a) real part of the electric field $E_r$, (b) real part of the electric field $E_\phi$, (c) imaginary part of the electric field $E_z$, (d) real part of the magnetic field $H_r$, (e) real part of the magnetic field $H_\phi$, (f) imaginary part of the magnetic field $H_z$.

The mode dispersion curves for the semicircular silver nanoridges of different radii are calculated and shown in Fig. 4. The black solid line is the light line in air, and the black dashed line is the dispersion curve of the metal-air flat surface plasmon mode. It can be seen that for all the radii, as the frequency increases, the mode dispersion curves shift away from the light line in air, suggesting slower group velocity and tighter mode confinement. This effect is far more pronounced for nanoridges with smaller radius. As the radius increases from 25 $nm$ to 500 $nm$, the dispersion curve moves toward the light line, indicating the reduction of the mode confinement and the propagation attenuation. As the radius of the semicircular ridge continues to increase, the ridge plasmon mode gradually approaches to the plasmon mode of the metal-air flat surface.

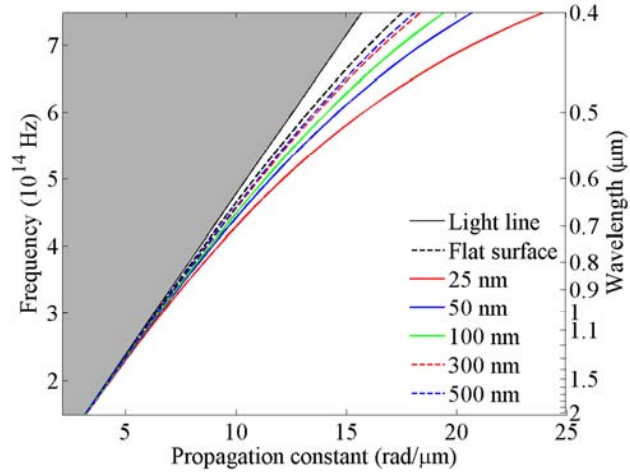

Fig. 4. Dispersion curves of the silver semicircular nanoridge plasmon waveguides of different radii and the comparison with the dispersion curve of the metal-air flat surface plasmon mode.

The real (solid) and imaginary (dashed) parts of the mode index of the semicircular nanoridge (black) and cylindrical plasmon waveguide (blue) versus the radius at 1.55 $\mu m$ wavelength are shown in Fig. 5. Interestingly, both the real and imaginary parts of the mode index are seen to decrease with the increase in the radius, indicating a reduction in propagation attenuation and mode confinement. The real and imaginary parts of the mode index of the semicircular nanoridge are much smaller than those of the cylindrical plasmon waveguide of the same radius. This indicates that the semicircular metal nanoridge plasmon mode has lower propagation attenuation than the plasmon mode of the cylindrical silver nanowire with the same radius.

In general, the propagation of the surface plasmon mode can be characterized by a complex wave propagation constant in the form of $\beta_z = n_{eff} k_0 = \beta - j\alpha$, along the $z$-direction, where $\beta$ is the phase propagation constant, and $\alpha$ is the attenuation constant. The propagation distance ($L_p$) is defined as the distance where the mode intensity attenuates to $1/e$ of its initial value, i.e. $L_p=1/(2\alpha)$. Outside the metal in the surrounding dielectric, the transverse component of the complex wave vector component is $\beta_\perp = \gamma - j\delta$, where $\gamma$ and $\delta$ describe the field oscillation and decay in the transverse direction, respectively. Following the Maxwell's equations, the complex wave mode propagation constant in the propagation direction and the complex wave vector component in the transverse direction are related as:

$$(\beta - j\alpha)^2 + (\gamma - j\delta)^2 = \varepsilon_d k_o^2 \qquad (1)$$

where $k_0$ is the mode propagation constant in the free space, and $\varepsilon_d$ is the dielectric constant of the surrounding dielectric. Solving (1), we can obtain $\gamma$ and $\delta$, which can be used to define the mode size as $1/2\delta+1/2\delta=1/\delta$.

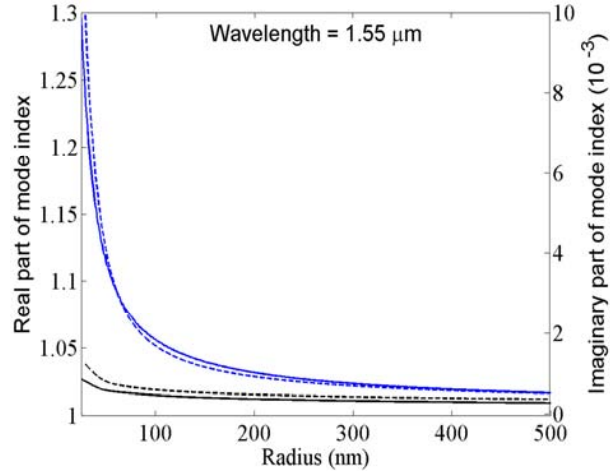

Fig. 5. Real (solid) and imaginary (dashed) parts of the mode index of the semicircle nanoridge (black) and cylindrical plasmon waveguide (blue) versus the radius at 1.55 $\mu m$ wavelength.

While it is always desirable to have optical waveguides with tight confinement and low propagation attenuation, the reality is that there is always a trade-off between the propagation attenuation and the mode confinement for surface plasmon waveguides [11, 52]. While tight mode confinement is the merit, large attenuation is the cost. Figure-of-merits of surface plasmon waveguides have been proposed to characterize this trade-off between the attenuation and the confinement [53, 54]. Intuitively, we define the figure-of-merit (FoM) of the nanoridge plasmon waveguide as the ratio of the propagation distance over the mode size:

$$FoM = (1/2\alpha)/(1/\delta) = \delta/2\alpha \qquad (2)$$

We calculated the propagation distance and the figure-of-merit of the semicircular silver nanoridge and the cylindrical silver plasmon waveguide versus the radius of the curvature at the wavelength of 1.55 $\mu m$. The results are shown in Fig. 6. Fig. 6 shows the propagation distance (solid) and the figure-of-merit (dashed) of the semicircle top silver nanoridge (black) and cylindrical silver nanowire (blue) versus the radius at 1.55 $\mu m$ wavelength. It can be seen that as the radius increases, both the propagation distance and figure-of-merit increase. The propagation distance and figure-of-merit of the semicircular silver nanoridge mode are larger than those of the cylindrical silver plasmon waveguide of the same radius.

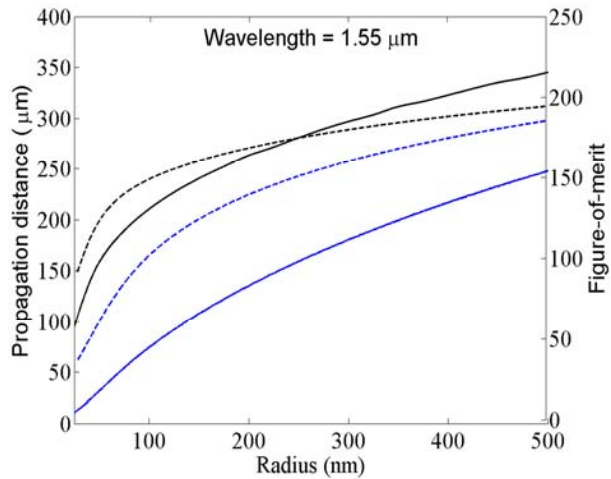

Fig. 6. Propagation distance (solid line curves) and figure-of-merit (dashed line curves) of the semicircle top silver nanoridge (black) and cylindrical silver nanowire (blue) versus the radius at 1.55 $\mu m$ wavelength.

The real and imaginary parts of the semicircular nanoridge mode index versus the wavelength for several different radii ($r$ = 25 *nm*, 50 *nm*, 100 *nm*, 300 *nm*, and 500 *nm*) are plotted in Fig. 7, where the black dashed lines are those of the metal-air flat surface plasmon mode. As the wavelength increases, both the real and imaginary parts of the semicircular nanoridge mode index decrease, indicating a reduction in the mode confinement as well as the propagation loss. It can also be seen that when the radius increases, both the real and imaginary parts of the semicircular nanoridge mode index reduce, which is consistent with the results shown in Figs. 2, 4, 5, and 6.

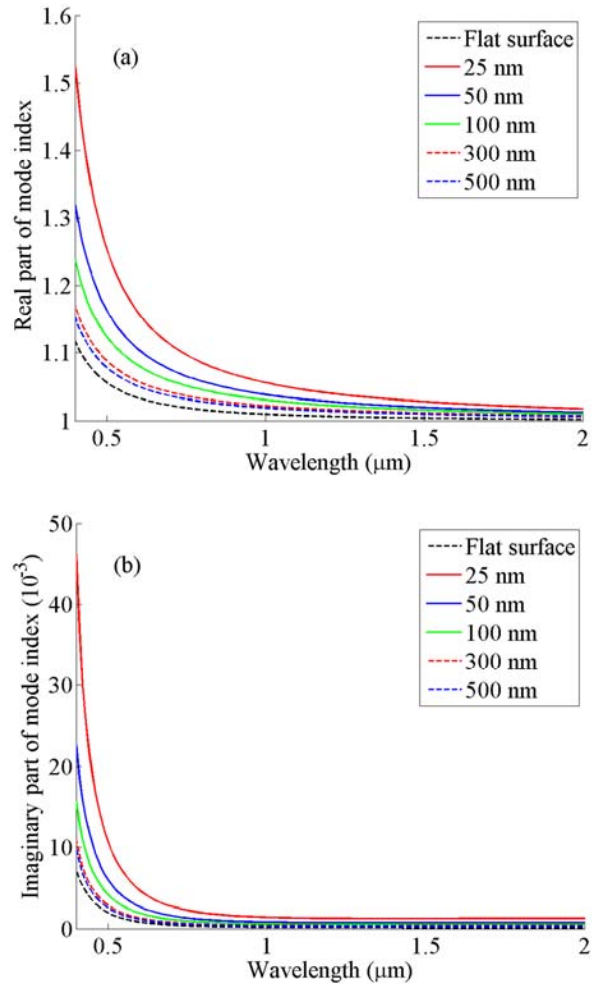

Fig. 7. (a) Real part and (b) imaginary part of the mode index of the semicircular nanoridge versus the wavelength for different radii and the comparison with those of the metal-air flat surface plasmon mode index.

The plasmon mode propagation distances are also calculated for the semicircular nanoridges of different radius at different free space wavelengths. The results are shown in Fig. 8 (a) and (b). It can be seen that as the wavelength or radius increases, the propagation distance increases.

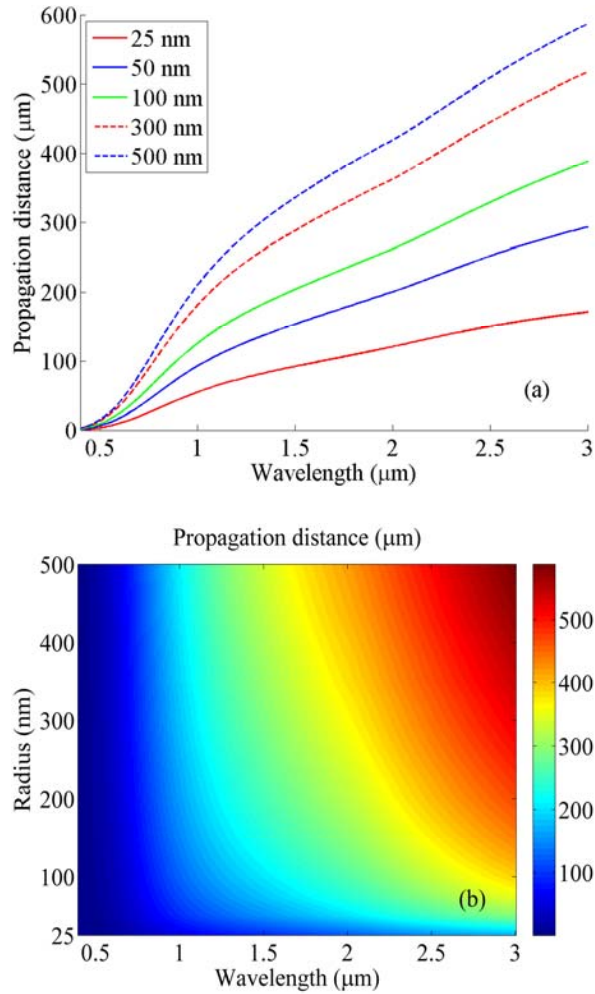

Fig. 8. (a) Propagation distance of the semicircular top nanoridge versus the free space wavelength for different radii. (b) Propagation distance of the semicircular top nanoridge mode versus the wavelength and the radius.

We also calculated the mode size versus the free space wavelength for different radii. The results are shown in Fig. 9. Fig. 9 (a) shows the mode size of the semicircular silver nanoridge versus the free space wavelength for several different radii; Fig. 9 (b) show the mode size of the semicircular nanoridge plasmon waveguide versus the wavelength and radius. It can be seen that as the mode size increases as the wavelength or radius increases.

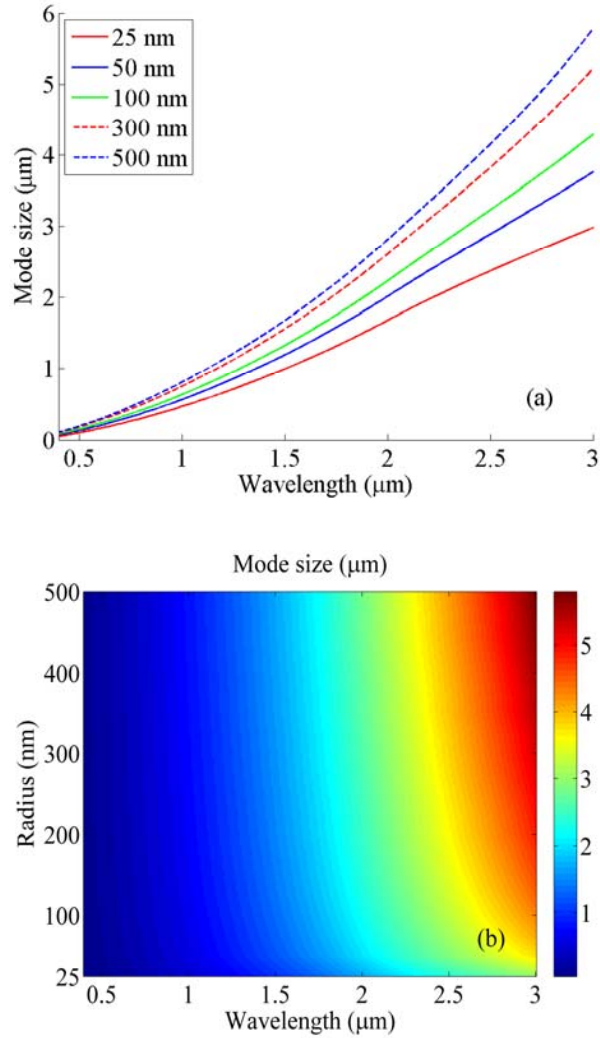

Fig. 9. (a) Mode size of the semicircular nanoridge plasmon versus the free space wavelength for several different radii. (b) Mode size of the semicircular nanoridge plasmon waveguide versus the wavelength and radius.

Here, we define a confinement factor (*CF*) for the nanoridge waveguide mode as the ratio of the wavelength over the mode size:

$$CF = \lambda/(1/\delta) = \lambda\delta \tag{3}$$

We calculated the confinement factor of the semicircular nanoridge waveguide versus the wavelength for several different radii. The results are shown in Fig. 10(a). Fig. 10(b) shows the confinement factor versus the wavelength and radius. It can be seen from Fig. 10(a) and (b) that as the radius or the wavelength increases, the confinement factor decreases. The strong mode confinement happens at the small radius of nanoridges and at the short wavelength.

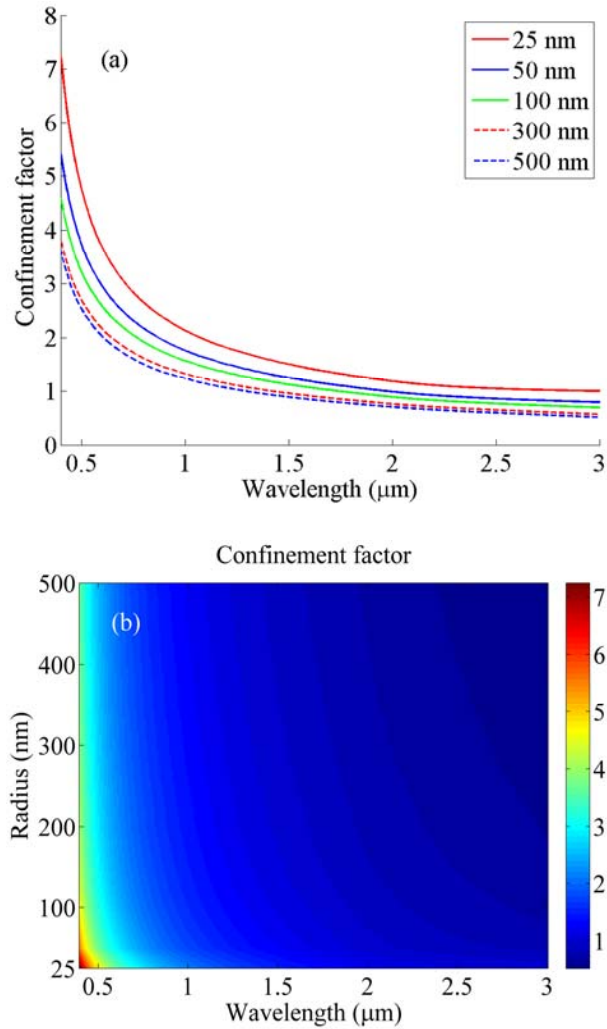

Fig. 10. (a) Confinement factor of the semicircular nanoridge plasmon mode versus the free space wavelength for different radii. (b) Confinement factor of the semicircular nanoridge mode versus the wavelength and radius.

We calculated the figure-of-merit of the semicircular nanoridge plasmon waveguide mode for different wavelengths and different radii of curvature. Fig. 11(a) shows the figure-of-merit versus the wavelength for several different radii. Fig. 11(b) shows the 2D plot of the figure-of-merit versus the wavelength and the radius. It can be seen from Fig. 11(a) and (b) that the figure-of-merit reaches a maximum at 1.05 $\mu m$ wavelength for all radii considered in this study. It is due to the optical properties of the silver metal and the surrounding dielectric material that the figure-of-merit peaks around the 1.05 $\mu m$ wavelength. This suggests 1.05 $\mu m$ as the optimal operational wavelength for silver nanoridge waveguides embedded in the air. It is also seen that at any wavelength, the figure-of-merit increases as the radius increases.

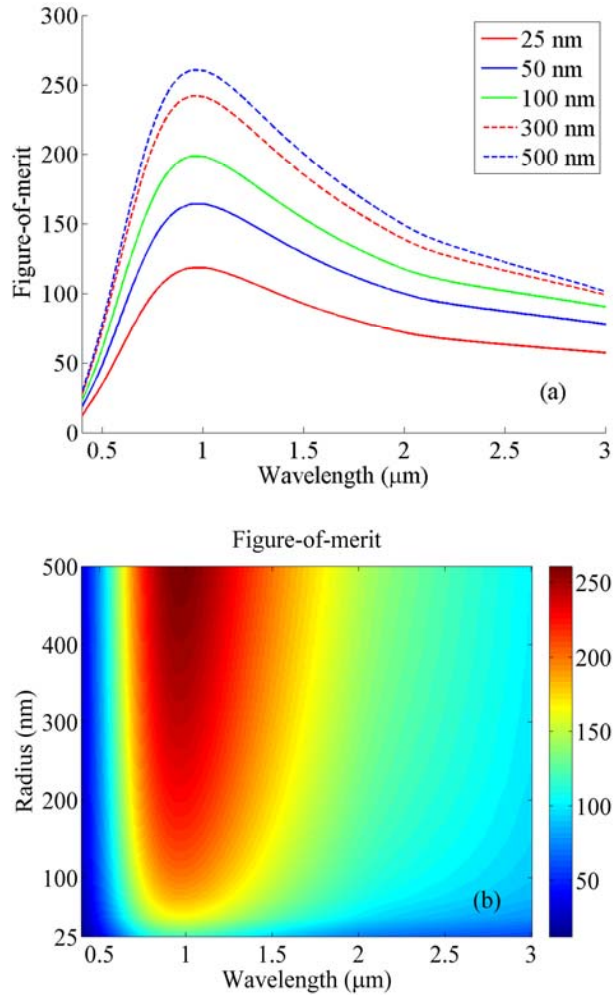

Fig. 11. (a) Figure-of-merit of the semicircular nanoridge versus the wavelength for several radii. (b) Figure-of-merit of the semicircular silver nanoridge plasmon mode versus the wavelength and the radius.

## 3. Summary

We investigated the guided plasmon modes supported by semicircular cross section top silver nanoridges. We calculated the mode field profiles, dispersion curves, propagation distances, mode sizes, confinement factors, and the figure-of-merits of the silver metal nanoridge plasmon waveguide modes. It is found that as the radius of the semicircular nanoridge increases, both the propagation distance and the figure-of-merit increase, but the confinement decreases dramatically. The semicircular top metal nanoridge waveguides have better performance in terms of the propagation distance and figure-of-merit than that of the cylindrical metal nanowires. The semicircular top metal nanoridges, which give more uniformly mode energy distribution over the entire ridge surface than the flat-top and triangular cross metal nanoridges [44, 45], represent an optimal trade-off between the mode confinement and propagation distance. Semicircular cross section top nanoridge plasmon waveguides investigated in this paper are very desirable for applications that require large surface interaction area, such as in chemical and biological sensing.

## Acknowledgment

This work was partially sponsored by the National Science Foundation (NSF) through the award NSF-0814103.